# Automatic Generation of Programming Exercises

Peter Sovietov
*Institute of Information Technologies*
MIREA – Russian technological university
Moscow, Russia
sovetov@mirea.ru

*Abstract*—Massive training of developers following the growing demands of the information technology industry requires teachers to automate their repetitive tasks. For training courses on programming, it is promising to use automatic generation and automatic grading of exercises that require a student to write a program. This article discusses the general scheme for constructing a programming exercises generator and identifies two classes of exercises, the generation of which can be automated: converting notation into code and converting data formats. Several examples of programming exercise generators are discussed. The experience of using exercise generators for the Python programming course is briefly described.

## I. Introduction

The rapid development of the information technology industry requires the coming in of more and more new developers. For the mass training of such specialists, universities use e-learning technologies. It raises the problem of ensuring the quality of education and is especially acute when organizing practical programming lessons.

The teacher regularly faces the need to create programming exercises both as part of the development of the new training courses and due to publishing existing material in the open access. No less significant are the problems of detailed checking of solutions and personalization of teaching, including the individual selection of the complexity of exercises and obtaining detailed feedback on errors.

The use of programming exercise generators can help solve the scalability problem of training courses. For wide practical use, exercise generators should have the following capabilities:

- Providing the necessary variety of exercises to solve the problem of plagiarism.
- Generating meaningful exercises that reflect significant concepts and programming techniques.
- Adaptation of the difficulty of the exercise to the level of a particular student.
- Automatic testing of the correctness of the student's solution.

The use of exercise generators in the educational process is nothing new. There are, in particular, generators of exercises in algebra [1], geometry [2, 3], mathematical logic [4], automata theory [5], as well as for the course of embedded systems [6]. Much less work is devoted to automating the generation of exercises that require a student to write a program as a solution.

In [7], a method for generating entry-level programming exercises is described. The generator builds a random program without loops, that consists of simple operations. Then, a text in natural language is generated, which describes the sequence of operations in the program. The student's task is to translate the natural language text into the chosen programming language.

The disadvantages of this generator are the following:

- A random program to be written does not belong to any applied domain, so the student may not be interested in the solution.
- The solution is trivial, without using the approaches on the different levels of the problem.

This article presents a general scheme for making programming exercise generators. These generators produce exercises from the various applied domains like data wrangling, bit-level programming, or binary data formats processing. Some of these generators use a hierarchical approach to build an exercise: the building blocks selected on the current level have an impact on the choosing of the elements of the lower levels. The practical use of described methods is shown in the example of the Python language training course at RTU MIREA University.

## II. General Scheme of a Programming Exercises Generator

In the proposed scheme shown in Fig. 1, exercise generation contains the following main steps:

1) Creation of a random exercise.
2) Creation of a test set.

A hash transformation of student's data is using as a seed for the exercise generator, which provides each student with an individual set of exercises. The difficulty level of the exercise is varied using input parameters of the exercise generator.

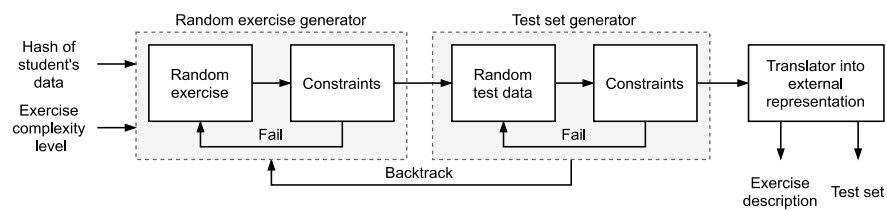

Fig. 1. General scheme for constructing a programming exercises generator.

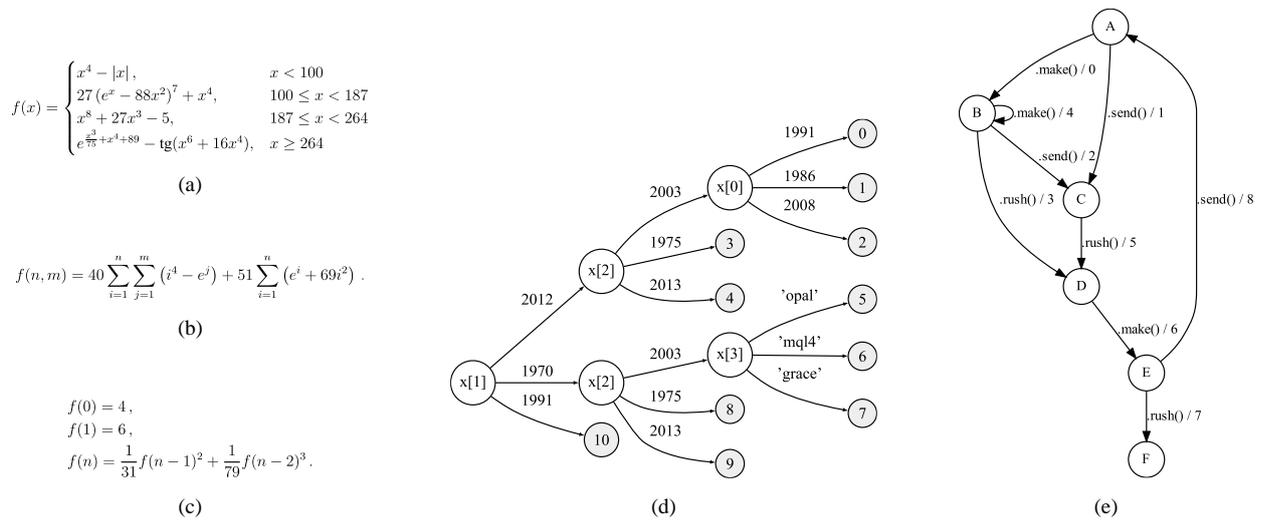

Fig. 2. Examples of fragments of exercises for converting a notation into a code. (a-c) Mathematical expression. (d) Decision tree. (e) Finite-state automaton.

It is possible to define the following classes of generated exercises:

- Converting notation to code. Examples of input notations: mathematical notation, decision tree, state machine, UML diagrams.
- Conversion of data formats. Examples: processing tabular data, parsing various binary and text data formats.

The random exercise generator contains two modules:

1) Creation of potential exercise.
2) Checking the result using a set of constraints.

Modularity simplifies implementation and allows the reuse of individual parts of the exercise generator.

If the generated random exercise has not satisfied the constraint set, then it is discarded. The result generated using the random exercise generator depends on the chosen class of exercises. It can be:

- A program in some intermediate representation.
- A description of the input and output data formats.

Depending on the exercise class, the test set generator may use an intermediate representation interpreter or a structured data generator.

When generating a test suite, the result is checked using the specified constraints, such as the uniqueness of test cases or the number of generated test cases.

If it was not possible to form a set of tests with the specified constraints, then backtracking is used to the start of the random exercise generator.

The test case is a pair $(x, y)$, and the form of the input and output depends on the type of code expected as a solution:

- Pure function. The input is a tuple of arguments, and the output is a tuple of results.

- Stateful code. Input data is a sequence of calls (methods, messages), and output data is a sequence of results.

If the test suite generation was successful, the internal representation of the result is translated into the external form, which may include text in a natural language or some visualization using tools like LaTeX, Graphviz, and PlantUML.

III. GENERATORS OF EXERCISES ON CONVERTING NOTATION TO CODE

For many exercise generators converting some notation into code, it is common to construct a random expression as the abstract syntax tree. The parameters that determine the complexity of generating exercise are used, such as the maximum tree depth and the maximum number of nodes of various types. It is possible to specify these constraints for individual subtrees too.

It is convenient to describe the expression for generation in an executable BNF-like form, using a system of

```
c ::= alt(c_1, ..., c_n, prob = [p_1, ..., p_n], end = c_e)
    | const(n, range = [n_1, n_2])
    | binop(add|sub|mul|div| ..., c_1, c_2)
    | ...
```

Fig. 3. Example of combinators to generate an abstract syntax tree.

combinators (higher-order functions for making domain-specific languages), see Fig. 2.

One of such combinators is $alt$ which randomly selects its argument for execution. This choice is made using a given probability distribution ($prob$).

Generating an expression by random depth-first search with $alt$ can lead to infinite recursion. There are ways to solve this problem [8]. For example, $alt$ may have an argument called $end$, which defines the combinator for the leaf of a tree. The use of this combinator happens by violating constraints on the depth or the number of generated nodes.

The generation of a random program may end with algebraic simplifications of the resulting expression tree.

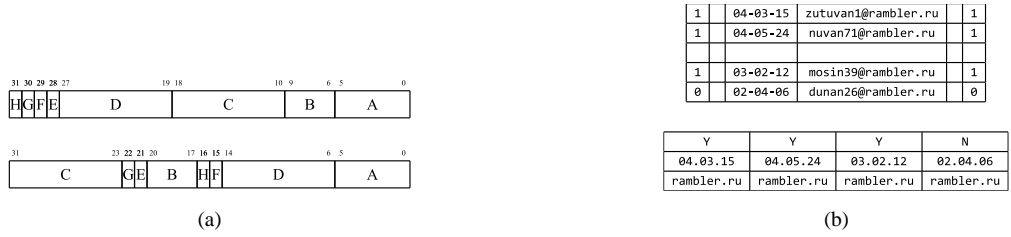

Fig. 4. Examples of fragments of exercises for converting data formats. (a) Permutation of bit fields. (b) Transformation of the table. (c) Parsing binary data.

When creating a set of tests, the interpreter function is used, which accepts the program and the environment (a table of variables and their test values).

Generators of exercises, see Fig. 3(a-c), are made to train basic skills in working with conditionals, loops, and recursion. Results are obtained using LaTeX.

The following generator, see Fig. 3(d), uses graphical notation — decision tree. The generation starts with choosing attributes and their types. Next, a decision tree is built, using the specified constraints on the attributes number, the range of attribute values, the depth of the tree, and the number of returned values. These parameters determine the complexity of the task. The result is obtained using Graphviz and TikZ.

The generator shown in Fig. 3(e) trains skills in making finite-state automata. Constraints on the number of nodes and edges of the graph determine the complexity of the exercise. The graph generated in two steps:

1) Making a random chain of nodes.
2) Adding random edges that satisfy the given constraints.

Here a test set is generating for the stateful program, so a trace of method calls generated using a random walk on the graph.

## IV. GENERATORS OF EXERCISES ON CONVERTING DATA FORMATS

In the exercise generators of this section, random data formats are generated instead of programs. The test set generator fills the data with random values according to the generated structure.

A fragment of an exercise that develops skills in working with bit fields is shown in Fig. 4(a). In a word of a given length, it is necessary to permute the given bit fields. Word size, maximum field size, and the maximum number of data fields are parameters for the complexity of the problem.

The fragment of the exercise shown in Fig. 4(b) corresponds to a typical task of converting data from one tabular format to another one. The complexity of this exercise depends on the number of columns in the table, the allowed data types, and the allowed transformations.

Generating the exercise under consideration includes creating a set of columns with the definition of their data types. Next, the permissible transformations of rows and columns are determined, which include:

- Split a column by a separator character.
- Delete empty rows and columns.
- Remove duplicates among rows and columns.
- Sort by a given column.
- Cells transformation by examples.
- Transpose table.

The transformation of cells occurs using a set of operations depending on the type of the column. Each of such operations uses constraints under which this operation is allowed. For example, the transformation of an area code in the "telephone number" type is only valid if there is an area code in the input data.

In the generator of exercises for parsing binary data, see Fig. 4(c), the hierarchical data format is generated that includes several base data types, a reference type, an array type, and a structure type. The complexity of the exercise depends on the maximum number of structure fields, the size of the arrays, and the depth of the generated data type tree. Building the type tree is similar to the building of the abstract syntax tree from the previous section.

Below is a set of constraints on the generation of a random binary data format:

$$(2 \leq D \leq 4) \wedge (40 \leq S \leq 160) \wedge (A \geq 4)$$
$$\wedge (R \geq 4) \wedge (A_r > 0)$$

This formula contains the maximum tree depth ($D$), the total data size in bytes ($S$), the number of arrays ($A$) and structures ($R$), and the arrays of structures ($A_r$) number.

## V. PRACTICAL USE OF PROGRAMMING EXERCISE GENERATORS

The programming exercise generators discussed in this article are implemented for the Python programming course at RTU MIREA University. Practical assignments had about 18,000 different automatically generated exercises.

The descriptions of the generated exercises include a few test cases. It helps the students to clarify the meaning of the exercise description and independently check their solutions. When an error occurs, the student sees only the output data of a specific unsuccessful test case.

Table I shows the performance of programming exercise generators implemented in Python. The performance test uses the following configuration: Intel Core i7-3770 CPU 3.40 GHz, 32 GB RAM, Python 3.9.2. The results show that the proposed scheme for constructing programming exercise generators is fast enough to be used online.

TABLE I. PERFORMANCE RESULTS OF PROGRAMMING EXERCISE GENERATORS

| Type of generated exercise | Generator performance (exercises/sec) |
|---|---|
| Translation of math notation | 17 |
| Implementation of decision tree | 125 |
| Permutation of bit fields | 6400 |
| Table data wrangling | 206 |
| Binary format parsing | 41 |
| Implementation of finite state automata | 12 |

In addition to automatic testing, implemented automatic hints [9–11] work as feedback for typical errors of students. The student's code (in the form of an abstract syntax tree or bytecode) is checked for the patterns previously written by the teachers.

Programming exercise generators have been used to test practical skills throughout the course and have helped provide the required variety of exercises for students. It is important to note that the automatically generated exercises were only an addition to the ones manually developed by the teachers.

## VI. CONCLUSION

This article examined the problem of scaling programming training courses and suggested using program exercise generators as one of the promising approaches to solving this problem.

A general scheme for constructing a generator of programming exercises is presented. Two main classes of exercises are identified, variations of which can be automated: exercises for converting notation into code and exercises for converting data formats.

The experience of using programming exercise generators in the Python language course shows that complex exercises can be generated online and individually for every student.

The approach proposed in the article demonstrates that programming exercise generators can produce a wide range of exercises and require non-trivial solutions from the student, thereby developing programming skills.

REFERENCES

[1] R. Singh, S. Gulwani, and S. Rajamani, "Automatically generating algebra problems," Proc. Conf. AAAI Artif. Intell., vol. 26, no. 1, 2012.

[2] C. Alvin, S. Gulwani, R. Majumdar, and S. Mukhopadhyay, "Automatic synthesis of geometry problems for an intelligent tutoring system," arXiv [cs.AI], 2015.

[3] C. Alvin, S. Gulwani, R. Majumdar, and S. Mukhopadhyay, "Synthesis of problems for shaded area geometry reasoning," in Lecture Notes in Computer Science, Cham: Springer International Publishing, 2017, pp. 455–458.

[4] U. Z. Ahmed, S. Gulwani, and A. Karkare, "Automatically generating problems and solutions for natural deduction," in Proceedings of the Twenty-Third international joint conference on Artificial Intelligence, 2013, pp. 1968–1975.

[5] P. Cerny, S. Gulwani, T. Henzinger, A. Radhakrishna, and D. Zufferey, "Specification, verification and synthesis for automata problems," technical report, 2012.

[6] D. Sadigh, S. A. Seshia, and M. Gupta, "Automating exercise generation: A step towards meeting the MOOC challenge for embedded systems," in Proceedings of the Workshop on Embedded and Cyber-Physical Systems Education - WESE '12, 2013.

[7] T. J. Tiam-Lee and K. Sumi, "Procedural generation of programming exercises with guides based on the student's emotion," in 2018 IEEE International Conference on Systems, Man, and Cybernetics (SMC), 2018, pp. 1465–1470.

[8] E. Soremekun, E. Pavese, N. Havrikov, L. Grunske, and A. Zeller, "Inputs from hell learning input distributions for grammar-based test generation," IEEE trans. softw. eng., pp. 1–1, 2020.

[9] R. Singh, S. Gulwani, and A. Solar-Lezama, "Automated feedback generation for introductory programming assignments," in Proceedings of the 34th ACM SIGPLAN conference on Programming language design and implementation - PLDI '13, 2013.

[10] S. Gulwani, I. Radiček, and F. Zuleger, "Feedback generation for performance problems in introductory programming assignments," in Proceedings of the 22nd ACM SIGSOFT International Symposium on Foundations of Software Engineering - FSE 2014, 2014.

[11] P. M. Phothilimthana and S. Sridhara, "High-coverage hint generation for massive courses: Do automated hints help CS1 students?," in Proceedings of the 2017 ACM Conference on Innovation and Technology in Computer Science Education, 2017.